\documentclass[aps,prb,preprint,groupedaddress]{revtex4}

\usepackage{graphics}
\usepackage{graphicx}
\begin{document}

\title{ Zero-temperature Phase Diagram For Strongly-Correlated Nanochains}
\author{Yan Luo, Claudio Verdozzi and Nicholas Kioussis}
\email[E-mail me at: ]{nick.kioussis@csun.edu. }
\affiliation{Department of Physics, California State University,
Northridge, California 91330-8268}
\begin{abstract}
{ Recently there has been a resurgence of intense experimental and
theoretical interest on the Kondo physics of nanoscopic and
mesoscopic systems due to the possibility of making experiments in
extremely small samples. We have carried out exact diagonalization
calculations to study the effect of the energy spacing $\Delta$ of
the conduction band on the ground-state properties of a dense
Anderson model nanochain. The calculations reveal for the first
time that the energy spacing tunes the interplay between the Kondo
and RKKY interactions, giving rise to a zero-temperature $\Delta$
versus hybridization phase diagram with regions of prevailing
Kondo or RKKY correlations, separated by a {\it free spins}
regime. This interplay may be relevant to experimental
realizations of small rings or quantum dots with tunable magnetic
properties.}

\end{abstract}
\maketitle

The possibility of making experiments in extremely small samples
has lead to a resurgence of both experimental and theoretical
interest of the physics of the interaction of magnetic impurities
in nanoscopic and mesoscopic non-magnetic metallic systems. A few
examples include quantum dots\cite{gordon}, quantum
boxes\cite{thimm} and quantum corrals\cite{manoharan}. Recent
scanning tunneling microscope(STM) experiments\cite{odom} studied
the interaction of magnetic impurities with the electrons of a
single-walled nanotube confined in one dimension. Interestingly,
in addition to the bulk Kondo resonance new subpeaks were found in
shortened carbon nanotubes, separated by about the average energy
spacing, $\Delta$, in the nanotube.  The relevance of small
strongly correlated systems to quantum computation requires
understanding how the infinite-size properties become modified at
the nanoscale, due to the  finite energy spacing $\Delta$ in the
conduction band \cite{thimm,Schlottman,Hu,Balseiro,Affleck}. For
such small systems, controlling $T_K$ upon varying $\Delta$ is
acquiring increasing importance since it allows to tune the
cluster magnetic behavior and to encode quantum information. While
the effect of $\Delta$ on the single-impurity Anderson or Kondo
model has received considerable
theoretical\cite{thimm,Schlottman,Hu,Balseiro,Affleck} and
experimental\cite{odom} attention recently, its role on {\it
dense} impurity clusters remains an unexplored area thus far. The
low-temperature behavior of a nanosized heavy-electron system was
recently studied within the mean-field
approximation\cite{Schlottman2}. A central question is what is the
effect of $\Delta$ on the interplay between the Kondo effect and
the RKKY interaction. The first interaction being responsible for
the quenching of the local $f$-moment (LM) via the screening of
the conduction electrons, whereas the latter being responsible for
magnetic ordering.

In this work we present exact diagonalization
calculations\cite{martin,dagotto} for $f$-electron nanochains
using periodic boundary conditions to study the effect of (1)
energy spacing, (2) $f$-electron conduction-electron hybridization
and (3) the parity of number of conduction electrons on the
interplay between the Kondo and RKKY interactions. While the
cluster properties depend on cluster geometry and
size\cite{Pastor}, the present calculations treat exactly the
Kondo and RKKY interactions.  Our results show that tuning
$\Delta$ and the {\it parity} of the total number of electrons can
drive the nanocluster from the Kondo to the RKKY regime, giving
rise to a zero-temperature energy spacing versus hybridization
phase diagram which is rich in structure.

We consider the half-filled ($N_{el}=2N)$ periodic Anderson
Hamiltonian for N=6 sites  arranged in a ring
\begin{eqnarray}
H &=& t\sum_{<ij>\sigma}c^{\dagger}_{i\sigma}c_{j\sigma}
+\sum_{i\sigma}\epsilon^i_{f}f^{\dagger}_{i\sigma}f_{i\sigma}
+\sum_{_i}U_{i}f^{\dagger}_{i+}f_{i+}f^{\dagger}_{i-}f_{i-}\nonumber\\
 & &+\sum_{i\sigma}V(f^{\dagger}_{i\sigma}c_{i\sigma}
+c^{\dagger}_{i\sigma}f_{i\sigma}).
\end{eqnarray}
Here, t is the nearest-neighbor hopping matrix element for the
conduction electrons, $c_{i,\sigma }^{+}(c_{i,\sigma })$ and
$f_{i,\sigma }^{+}(f_{i,\sigma })$ create (annihilate) Wannier
electrons in $c$- and $f$- like orbitals on site i with spin
$\sigma $, respectively. $E_{f}$  is the energy levels of the bare
localized orbital, V is the on-site hybridization matrix element
between the f and conduction orbitals, and U is the on-site
Coulomb repulsion of the f electrons. In this paper we consider a
simple tight-binding conduction band dispersion $
\epsilon_{k}=-2tcosk$ and the symmetric case $E_{f}=-\frac{U}{2}$,
with $U=5$.

We have investigated the ground-state properties as a function of
the hybridization and the energy spacing in the conduction band,
$\Delta = 4t/(N-1) = \frac{4t}{5}$. We have calculated the
 average $f-$ and $c-$local moments, $<(\mu_i^f)^2>$
and $<(\mu_i^c)^2>$, and the zero-temperature {\it f-f} and {\it
f-c} spin correlations functions (SCF) $<S_i^fS_{i+1}^f> \equiv
<g|S^{z,f}_iS^{z,f}_{i+1}|g>$ and $<S_i^fS_i^c> \equiv
<g|S^{z,f}_iS^{z,c}_i|g>$, respectively. Here, $|g>$ is the
many-body ground state and $S^{z,f}_i$ is the z-component of the
f-spin at site i. As expected, the cluster has a singlet ground
state ($S_g=0$ where $S_g$ is the ground-state spin). We compare
the  onsite Kondo correlation function $<S_i^fS_i^c>$ and the
nearest-neighbor RKKY correlation function $<S_i^fS_{i+1}^f>$ to
assign a state to the Kondo or RKKY regimes, in analogy with mean
field treatments\cite{lacroix}.

In Fig. 1 we present the variation of the local Kondo SCF
$<S_i^fS_i^c>$ (squares) and the nearest-neighbor RKKY SCF
$<S_i^fS_{i+1}^f>$ (circles) as a function of hybridization for
two values of the hopping matrix element $t=0.2$ (closed symbols)
and $t=1.2$ (open symbols), respectively. As expected, for weak
hybridization V the local nearest-neighbor RKKY (Kondo) SCF is
large (small), indicating strong short-range antiferromagnetic
coupling between the
 $f-f$ local moments, which leads to long range magnetic ordering for
 extended systems. As V increases,
$<S_i^fS_{i+1}^f>$ decreases whereas the $<S_i^fS_i^c>$ increases
(in absolute value) saturating at large values of V. This gives
rise to the condensation of independent local Kondo singlets at
low temperatures, i.e., a disordered spin liquid phase.
Interestingly, as $t$ or $\Delta$ decreases the {\it f-c} spin
correlation function is dramatically enhanced while the {\it f-f}
correlation function becomes weaker, indicating a transition from
the RKKY to the Kondo regime.

In Fig. 2 we present the average local $f$- (circles) and $c$-
(squares)  moments
 as a function of hybridization for
two values of the hopping matrix element $t=0.2$ (closed symbols)
and $t=1.2$ (open symbols), respectively.
 In the
weak hybridization limit, the large on-site Coulomb repulsion
reduces the double occupancy of the f level and a well-defined
local f moment is formed $\langle \mu_{f}^{2} \rangle = 1.0$ while
$\langle \mu_{c}^{2} \rangle = 0.5$. With increasing V both
charge- and spin- f1uctuations become enhanced and the local $f-$
moment decreases monotonically whereas the $c-$ local moment
exhibits a maximum. In the large V limit both the $f-$ and $c-$
local moments have similar behavior with $<\mu_{c}^{2}> \approx
\mu_{f}^{2} \approx \frac{1}{2}$ , indicating that the {\it total}
local moment $\mu$ vanishes. The effect of lowering the energy
spacing $\Delta$ is to decrease (increase) the $f-$ ($c-$) local
moment, thus tuning the magnetic behavior of the system. Note that
the
 maximum value of the $c-$ local moment increases as $\Delta$ decreases.
This is due to the fact that for smaller $t$ values the kinetic
energy of conduction electrons is lowered, allowing conduction
electrons to be captured by f electrons to screen the local f
moment, thus leading to an enhancement of the local $c-$ moment.

In Fig. 3 we present the energy spacing versus V zero-temperature
phase diagram of the nanocluster, which illustrates the interplay
between Kondo and RKKY interactions. In the RKKY region
$<S_i^fS_{i+1}^f>$ is larger than the $<S_i^fS_i^c>$ and the {\it
total} local moment is non zero; in the Kondo regime
$<S_i^fS_{i+1}^f>$ is smaller than the $<S_i^fS_i^c>$, the {\it
total} local moment vanishes, and the ground state of the system
is composed of independent local singlets. The red curve indicates
the crossover point, i.e., $<S_i^fS_{i+1}^f> = <S_i^fS_i^c>$. The
blue dashed curve denotes the set of points where the on-site {\it
total} local moment $\mu = 0$. Thus, in the intermediate regime,
which will be referred to as the {\it free spins} regime
\cite{coleman}, $<S_i^fS_{i+1}^f>$ is smaller than the $
<S_i^fS_i^c>$, the $f$ moment is {\it partially} quenched and $\mu
\not= 0$. Interestingly, we find that
 the {\it free spins} regime becomes narrower as
the average level spacing $\Delta$ is reduced. This result may be
interpreted as a quantum critical regime for the nanochain due to
the finite energy spacing, which eventually reduces to a quantum
critical point when $\Delta \rightarrow 0$.

We have also examined the effect  of changing $N_{el}$ from
$N_{el}$ = 12 ($S_g$ = 0) to $N_{el}$ = 11 ($S_g=\frac{1}{2}$) for
$t=1$. We find: (a) an enhancement of the local Kondo  SCF
$<S_i^fS_i^c>$ from  -0.01 to -0.12; and (b) a suppression of the
f-f SCF $<S_i^fS_{i+1}^f>$ from -0.58 to  -0.20 (due to the broken
symmetry for N$_{el}$ = 11, the f-f SCF's range from -0.5 to
+0.02). This interesting novel tuning of the magnetic behavior can
be understood in terms of the (single versus double) topmost
occupied conduction level: For $N_{el}$ even, double occupancy
prevents spin-flip transitions, thus weakening the Kondo
correlations.\cite{thimm}

In conclusion, we have carried exact diagonalization calculations
which reveal for the first time that the: (1) energy spacing;  and
(2) parity of $N_{el}$ give rise to a novel tuning of the magnetic
behavior of a {\it dense} Kondo nanochains. This interesting and
important tuning can drive the nanocluster from the Kondo to the
RKKY regime, i.e. a tunable $\Delta$ verus V zero-temperature
phase diagram at the nanoscale. The results indicate the presence
of an intermediate {\it free spins} regime which becomes narrower
as the energy spacing is reduced. Our conclusions should be
relevant to experimental realizations\cite{odom} of small clusters
and quantum dots, with appropriate tuning of the energy spacing.

The research at California State University Northridge (CSUN) was
supported through NSF under Grant Nos. DMR-0097187, NASA under
grant No. NCC5-513, and the Keck and Parsons Foundations grants.
The calculations were performed on the the CSUN Massively Parallel
Computer Platform supported through NSF under Grand No.
DMR-0011656.

\newpage

\begin{figure}[p]
\centerline{\includegraphics[width=.7\textwidth]{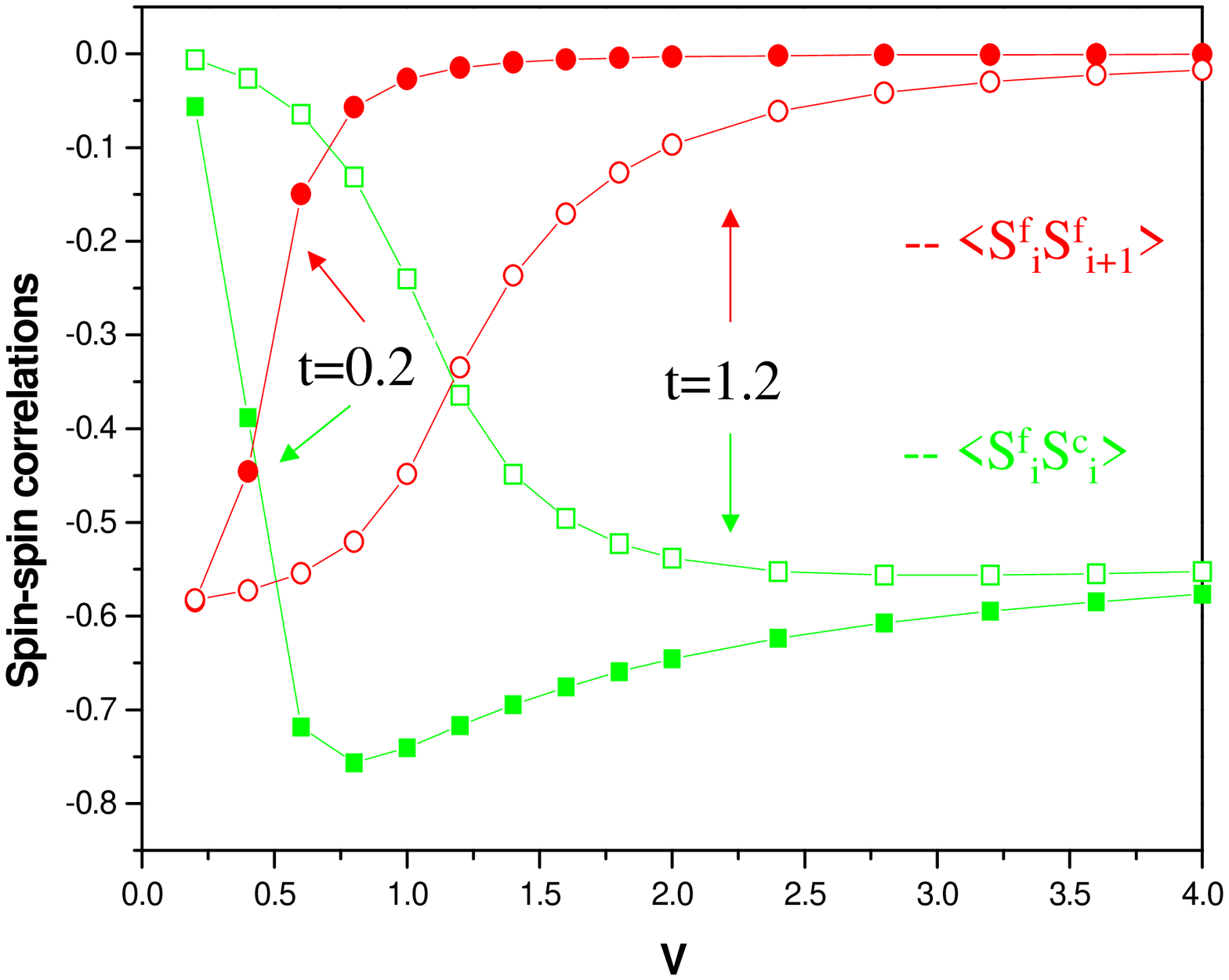}}
\caption{Nearest neighbor f-f spin-spin correlations (circles) and
on-site f-c spin-spin correlations (squares) as a function of V
for two values of the hopping parameter of $t=0.2$ (closed
symbols) and $t=1.2$ (open symbols), respectively.} \label{fig1}
\end{figure}

\newpage
\begin{figure}[p]
\centerline{\includegraphics[width=.7\textwidth]{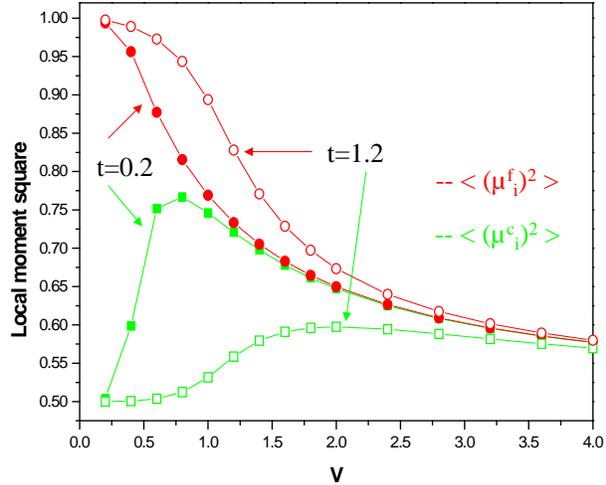}}
\caption{$f$- (circles) and $c-$ (squares) local moment  versus
hybridization for
 two  values of the hopping
parameter of  t=0.2 (closed symbols) and t=1.2 (open symbols),
respectively.} \label{fig2}
\end{figure}
\newpage
\begin{figure}[p]
\centerline{\includegraphics[width=.7\textwidth]{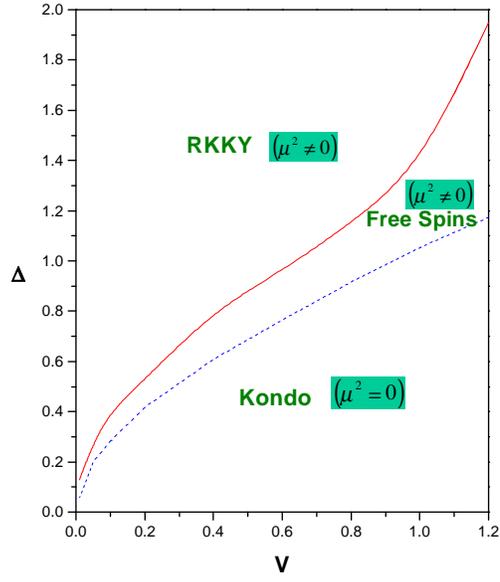}}
\caption{Energy spacing $\Delta$ versus hybridization
zero-temperature phase diagram. The red solid curve denotes the
crossover point of the spin-spin correlation function in Fig.1;
the blue dashed curve denotes the set of points where the on-site
{\it total} moment square $\langle (\mu_{f}+\mu_{c})^{2} \rangle =
0.0 \pm 0.05$.} \label{fig3}
\end{figure}

\end{document}